\documentclass[a4paper]{jpconf}

\usepackage{graphicx}
\usepackage{url}

\usepackage{enumitem}

\begin{document}

\title{Novel photon timing techniques applied to the LHCb RICH upgrade programme}

\author{Floris Keizer on behalf of the LHCb RICH collaboration}

\address{CERN, 1211 Meyrin, Switzerland}

\ead{floris.keizer@cern.ch}

\begin{abstract}
The Ring-Imaging Cherenkov (RICH) detectors at LHCb have an intrinsic time resolution of better than 10\,ps owing to the prompt Cherenkov radiation and focusing mirrors optics. While only spatial information has been used in the experiment to date, the addition of photon time information is one of the cornerstones of the future RICH upgrade programme. The novel timing techniques provide a powerful tool for background suppression and particle ID performance improvements. Here, developments to implement fast-timing in the front-end electronics are presented.
\end{abstract}

\vspace{-0.8cm}
\section{Introduction}

\begin{figure}[b!]
	\vspace{-0.6cm}
	\begin{center}
		\includegraphics[width=0.9\linewidth]{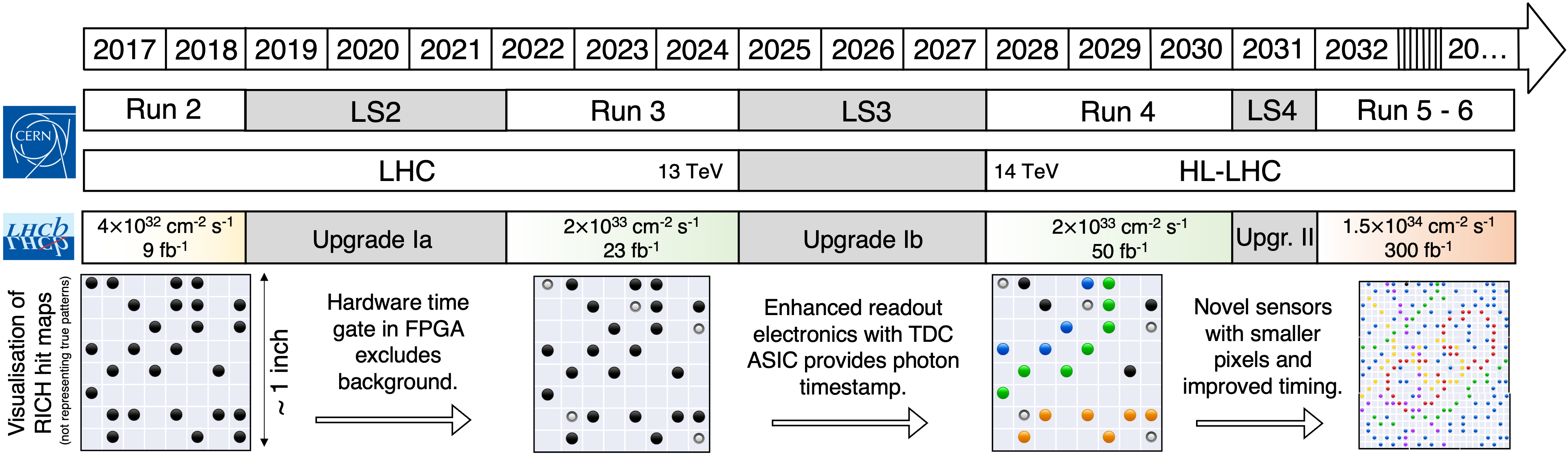}
	\end{center}
	\vspace{-0.7cm}
	\caption{\label{fig:timeline}Overview of the proposed upgrades of the RICH photon detector.
	% including time information with increased resolution.
	}
\end{figure}

\begin{figure}[t]
	\centering
	\begin{minipage}[t!]{.4\textwidth}
		\centering
		\includegraphics[width=0.77\textwidth]{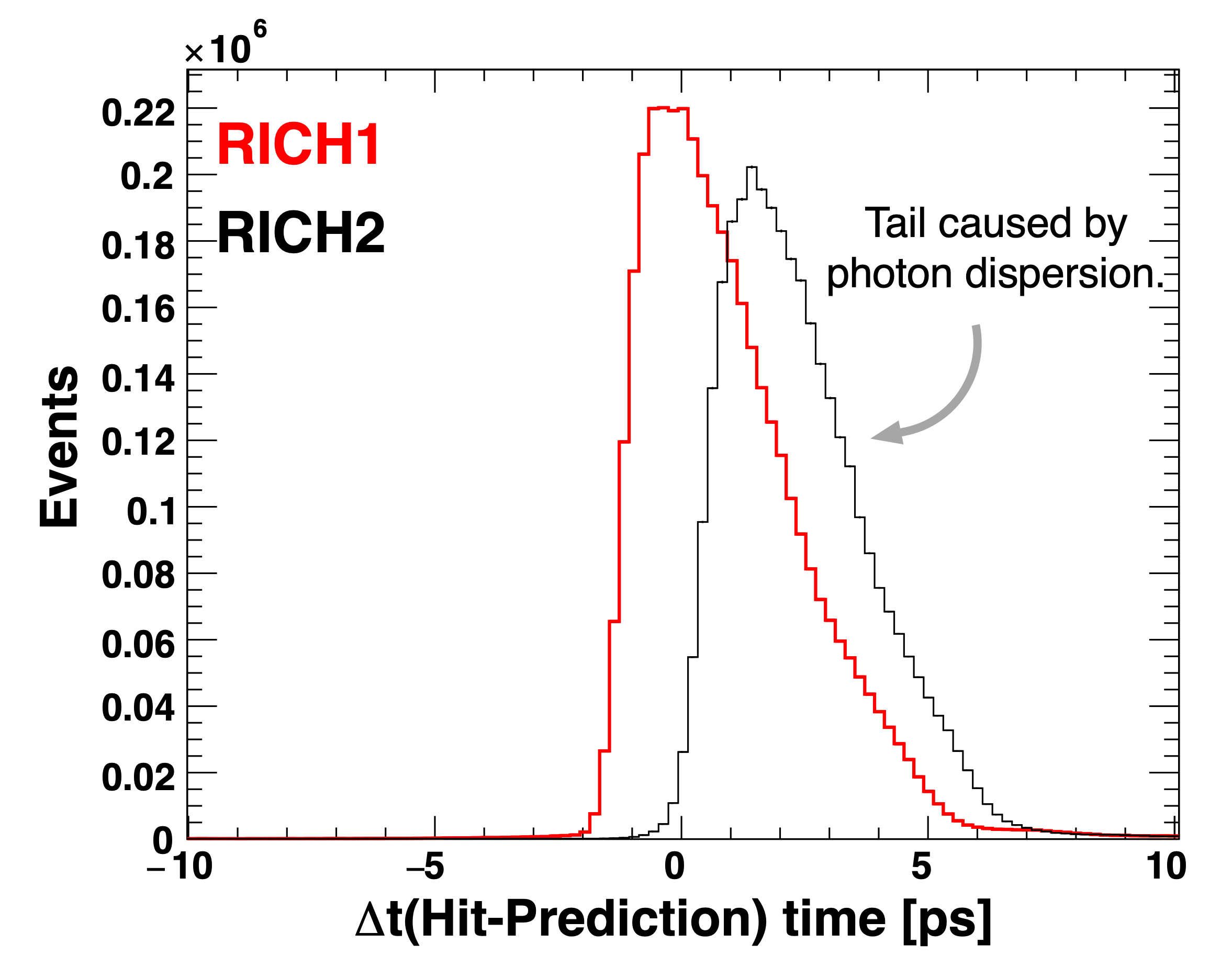}
		\vspace{-0.4cm}
		\caption{\label{fig:R1R2-ToA-res} Difference between the Cherenkov photon time of arrival and the predicted time using the RICH reconstruction algorithms.}
	\end{minipage}
	\hfill
	\begin{minipage}[t!]{.55\textwidth}
		\centering
		\includegraphics[width=1.0\textwidth]{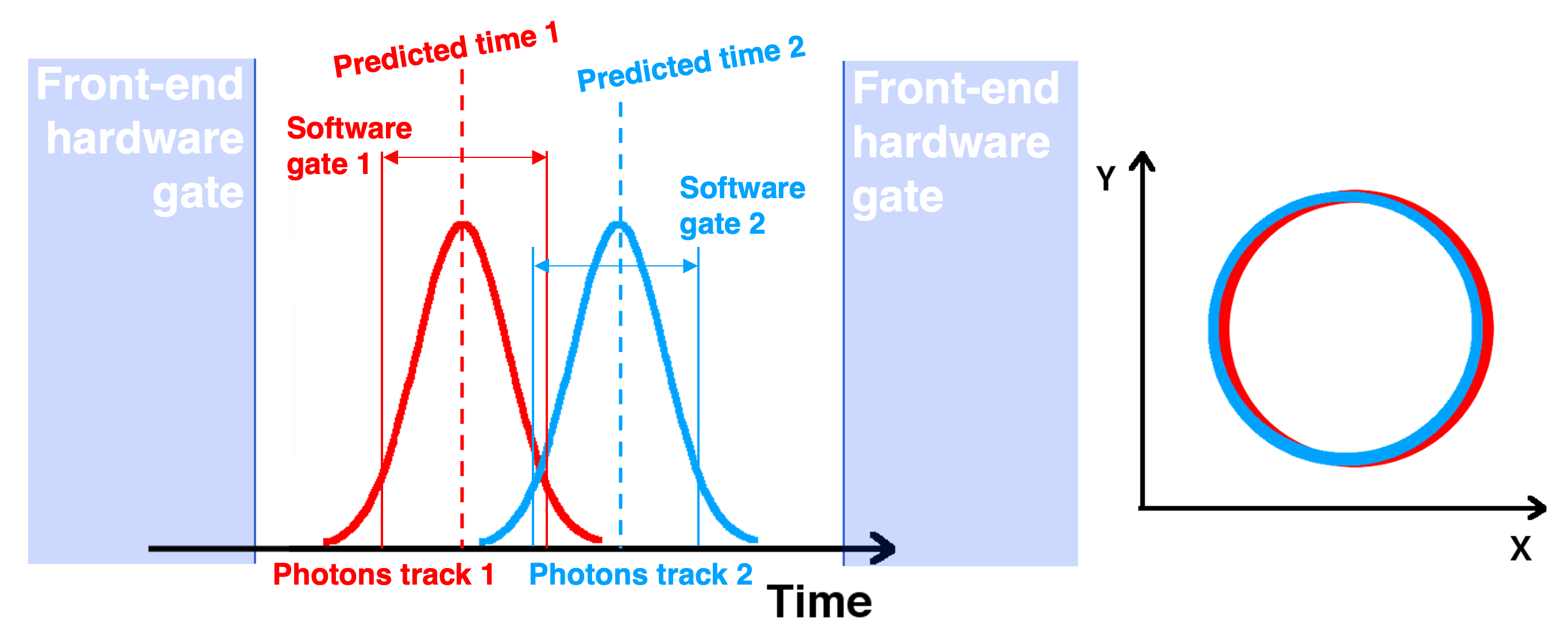}
		\vspace{-0.8cm}
		\caption{\label{fig:time-concept}Photon detector hit time information provides a novel solution to distinguish the signal from different tracks that overlap in the spatial domain.}
	\end{minipage} 
	\vspace{-0.5cm}
\end{figure}

\begin{figure}[b]
	\vspace{-0.6cm}
	\centering
	\begin{minipage}[t!]{.475\textwidth}
		\centering
		\includegraphics[width=0.75\textwidth]{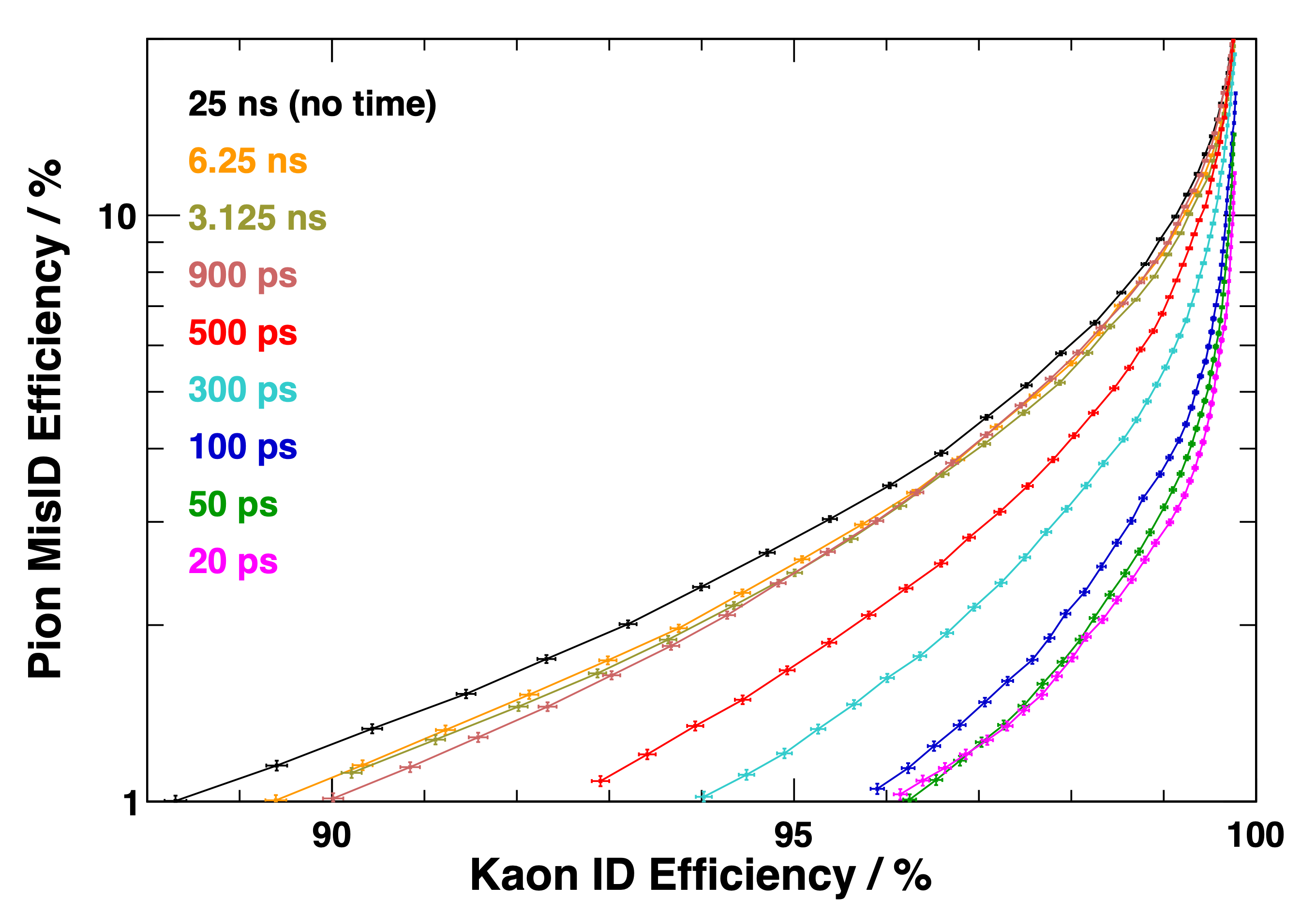}
		\vspace{-0.4cm}
		\caption{\label{fig:PID}Particle ID performance for different time gates around the predicted hit time in the Run~3 detector configuration. The absolute numbers depend on various detector optimisations, but the trend is consistent.}
	\end{minipage} 
	\hfill
	\begin{minipage}[t!]{.475\textwidth}
		\centering
		\includegraphics[width=0.7\textwidth]{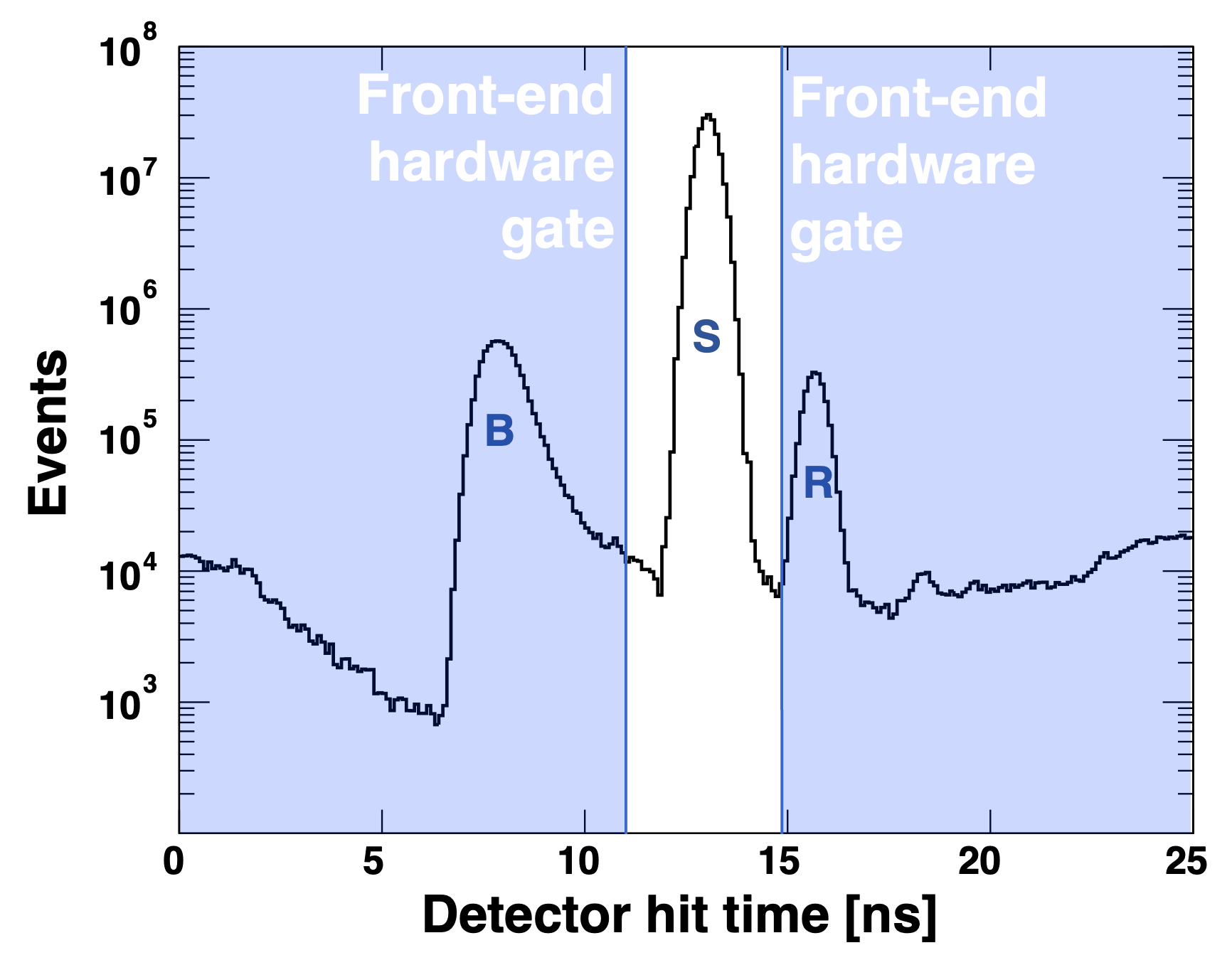}
		\vspace{-0.4cm}
		\caption{\label{fig:HW-gate}Simulated RICH detector hit time distribution showing the signal peak (S) and the backgrounds that would be eliminated using a FE hardware time gate.}
	\end{minipage}
\end{figure}

The LHCb upgrade programme aims to develop a novel time-resolved RICH detector with increasingly precise photon detector time resolution as outlined in Figure~\ref{fig:timeline}. This new avenue relies on the fact that for a given track the time-of-arrival (ToA) of Cherenkov photons on the detector plane can be predicted with an accuracy better than 10\,ps as shown in Figure~\ref{fig:R1R2-ToA-res} and has been demonstrated using the LHCb simulation framework \cite{Keizer-thesis}. The analytic prediction of the ToA in the RICH pattern recognition algorithms takes into account the LHCb tracking information, the track curvature in the LHCb magnetic field, the reconstructed photon paths in the RICH detector and the primary vertex (PV) time. For each track, a time gate can be applied in software around the predicted hit time, and only the hits arriving within this time gate are used for particle ID on that track. As illustrated in Figure~\ref{fig:time-concept}, this significantly reduces combinatorial background in particular from photons originating from one PV being associated with tracks originating from another PV. This is a powerful tool to improve the particle ID performance. Since the width of the software time gate is dictated by the time spread of the photon sensor and front-end (FE) electronics, these concepts motivate the RICH community to seek fast-timing technologies for the transition to the HL-LHC. A time gate of $\pm2\sigma_{sensor}$ was found to give a signal-to-background ratio that resulted in the optimal performance in simulation. The particle ID curves in Figure~\ref{fig:PID} provide a standardised and sensitive probe of the performance of the RICH log-likelihood optimisation. The curves show the trend of improved particle ID (the bottom-right corner represents a perfect performance) as software time gates of smaller width are applied around the predicted hit time. Here, sensor spread is not included in order to show the high potential of this timing technique using future technologies. \par

An important input to the hit time prediction is the PV time. Using the proposed readout enhancements, this can be estimated in the RICH algorithms during Run~4 using the multitude of photons from a PV combined with oversampling in the digital electronics to reduce the effect of sensor time spread. Preliminary  simulation studies demonstrate a resolution better than 100\,ps at least for a subset of PVs. This would be a new measurement for the RICH detector and LHCb experiment. During Run~5, the PV time is expected to be available from the tracking system with a picosecond resolution owing to the multitude of tracks per PV. \par

\vspace{-0.2cm}
\section{Detector front-end requirements}

The novel timing concepts set two requirements for the front-end electronics:
\begin{enumerate}[label=(\alph*)]
 	\vspace{-2mm}
	\item A nanosecond front-end hardware time gate.
 	\vspace{-1mm}
	\item Within this time gate, a timestamp for each detected hit.
 	\vspace{-1mm}
\end{enumerate}
The hardware time gate (a) acts like a shutter with a fixed, configurable latency with respect to the 40\,MHz LHC clock. The photon detector hit time distribution in Figure~\ref{fig:HW-gate} shows a signal peak (labelled S) as well as background including peaks from particles and photons travelling directly at the photon detectors (B) and photons undergoing additional reflections in the mirror system (R). The spread of the signal peak is dominated by the PV distribution and requires a time gate of approximately 2\,ns, which eliminates out-of-time background hits from the photon sensor and beam interactions. The hit timestamp (b) requires a time-to-digital converter (TDC) with bin size smaller than the sensor time resolution.\par

\vspace{-0.2cm}
\section{Nanosecond time gate during LHC Run 3}

\begin{figure}[t]
	\centering
	\begin{minipage}[t!]{.55\textwidth}
		\centering
		\includegraphics[width=1.0\textwidth]{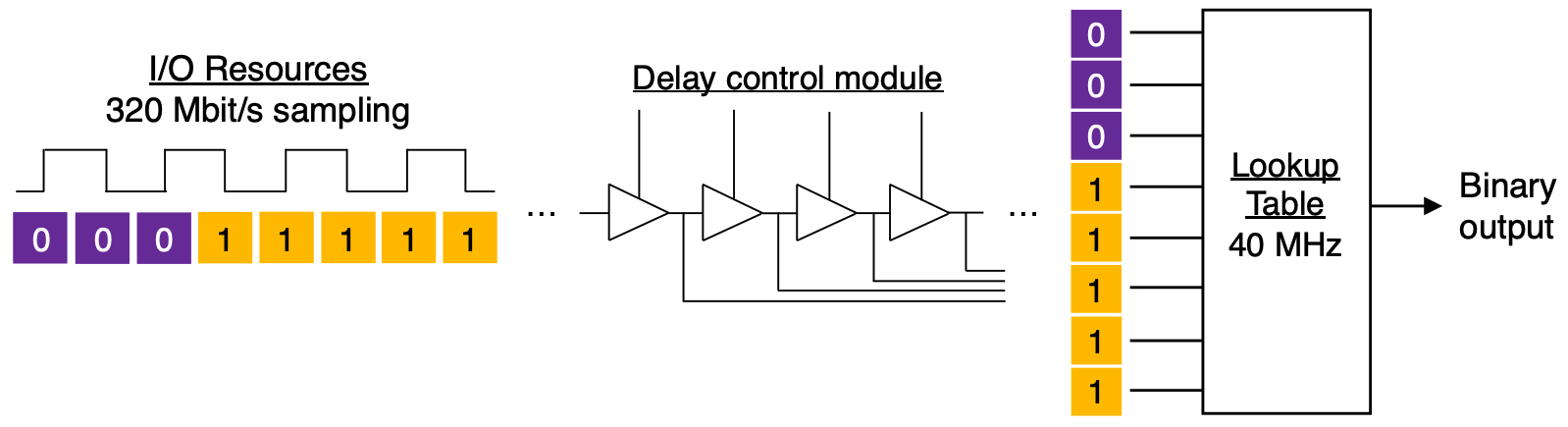}
		\vspace{-0.8cm}
		\caption{\label{fig:IO-logic}Schematic showing the time gate logic in the FPGA-based FE digital board during Run~3. The signals from the CLARO discriminator are sampled at 320\,Mbit/s and used to address a lookup table configured to accept specific input patterns.}
	\end{minipage}
	\hfill
	\begin{minipage}[t!]{.4\textwidth}
		\centering
		\includegraphics[width=0.7\textwidth]{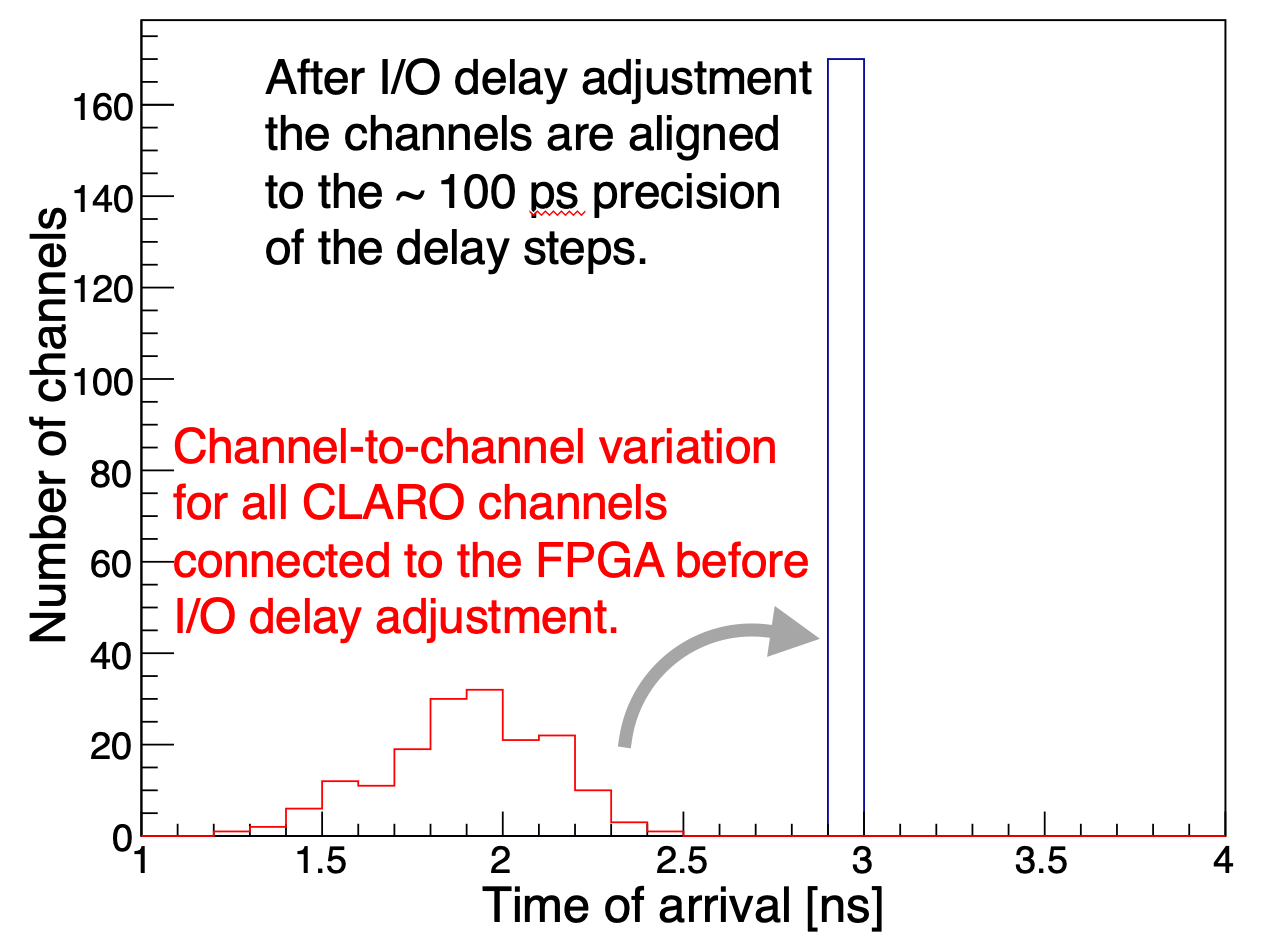}
		\vspace{-0.4cm}
		\caption{\label{fig:FPGA-calib}Channel-to-channel variation in the signals from the CLARO ASICs measured before (red) and after (blue) the delay control module.}
	\end{minipage} 
	\vspace{-0.4cm}
\end{figure}

\begin{figure}[b]
	\vspace{-0.4cm}
	\centering
	\begin{minipage}[t!]{.475\textwidth}
		\centering
		\includegraphics[width=0.7\textwidth]{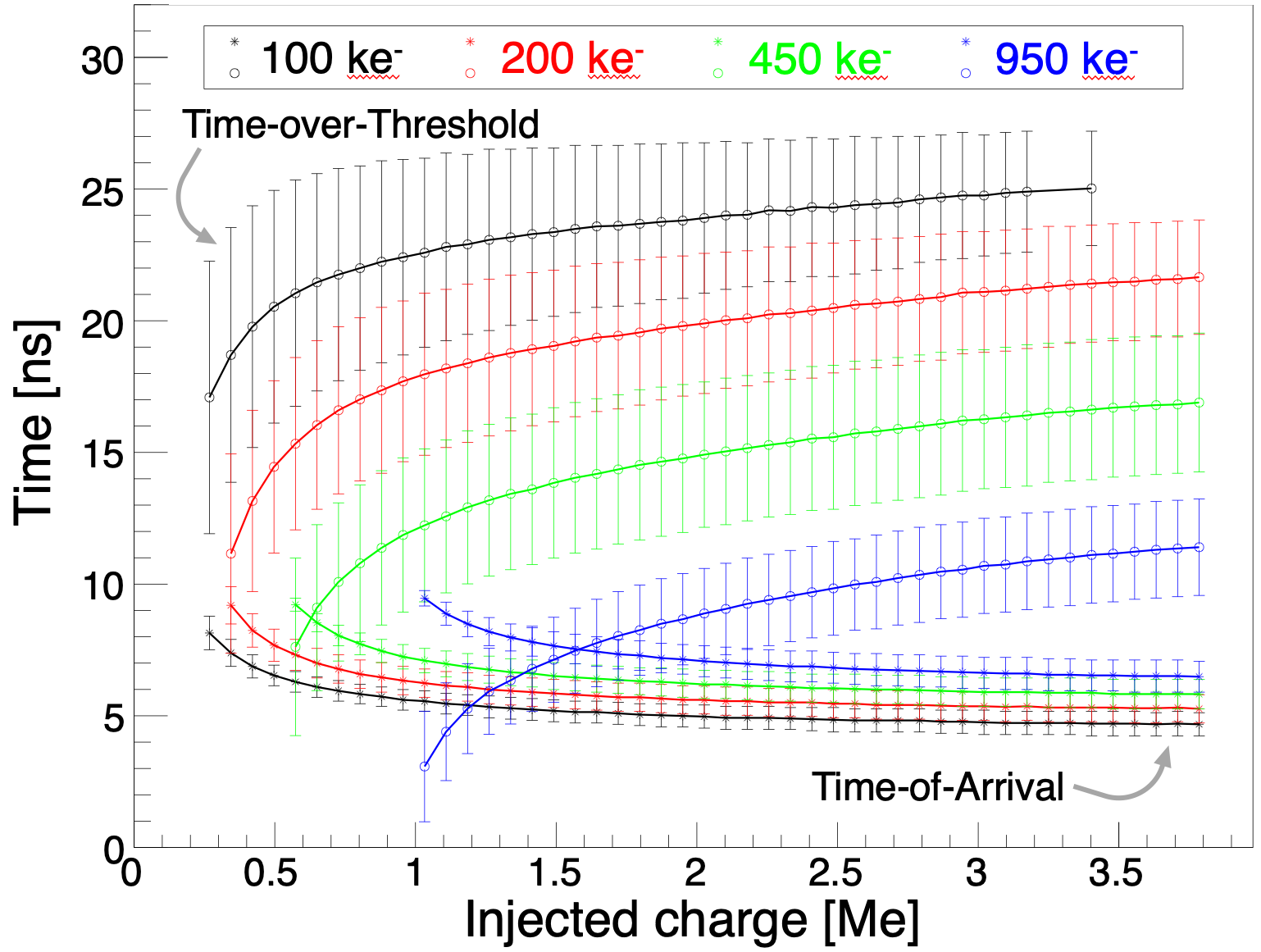}
		\vspace{-0.4cm}
		\caption{\label{fig:CLARO-toa-tot}Time-of-arrival and time-over-threshold for a range of injected test charge amplitudes measured at four different CLARO discriminator levels.}
	\end{minipage}
	\hfill
	\begin{minipage}[t!]{.475\textwidth}
		\centering
		\includegraphics[width=0.75\textwidth]{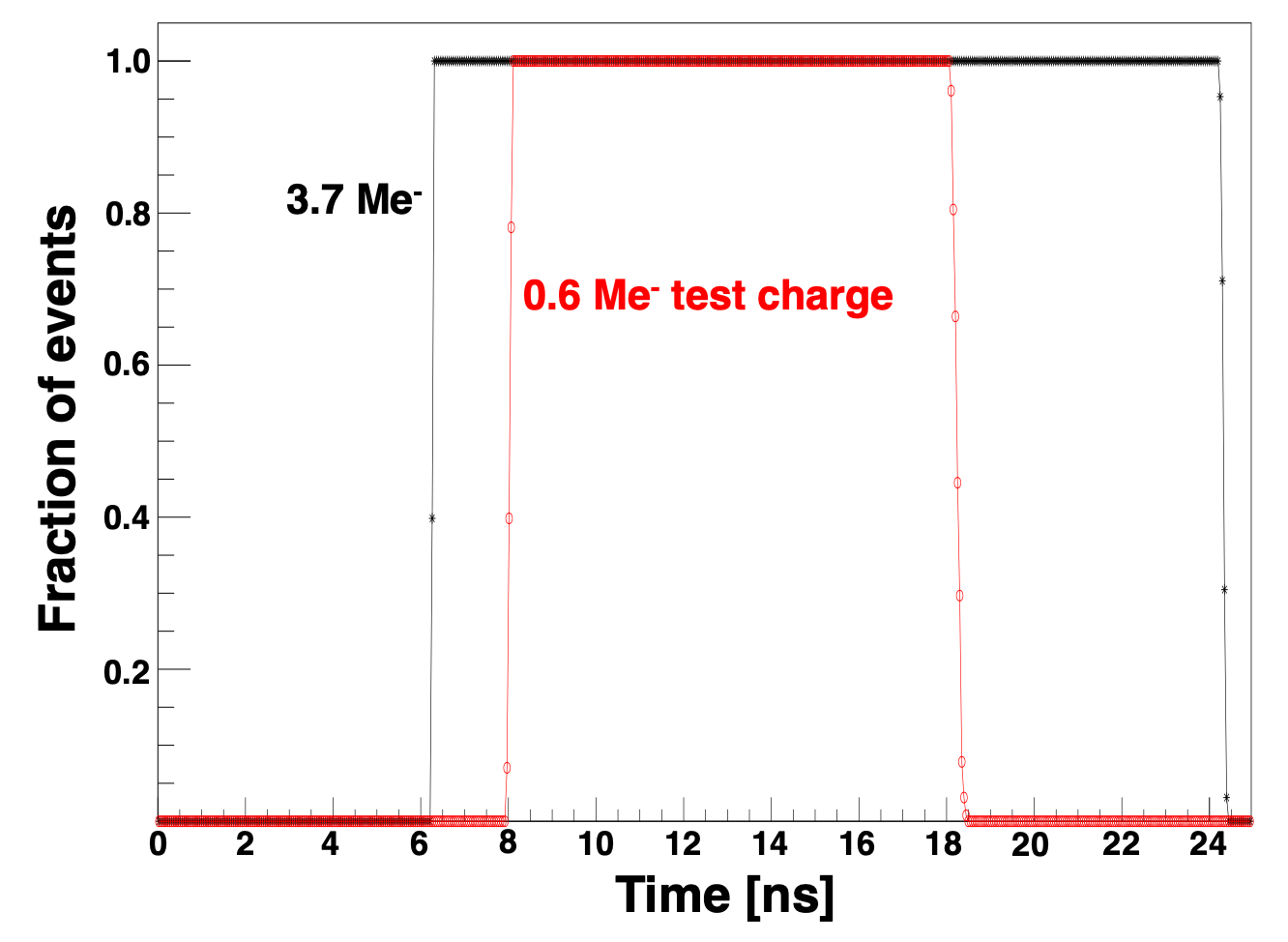}
		\vspace{-0.4cm}
		\caption{\label{fig:testpulse-timewalk}Digital signal from the CLARO ASIC captured by the FPGA for two test charge amplitudes, obtained by shifting the sampling clock phase.}
	\end{minipage} 
\end{figure}

The RICH photon detector uses multi-anode photomultiplier tubes (MAPMTs) during Run~3~and~4 \cite{TDR}. As outlined in Figure~\ref{fig:electronics-overview}, the MAPMT signals are shaped and discriminated by the CLARO readout ASICs \cite{CLARO}. The digital board provides the interface between the CLAROs and the data links to the LHCb readout. The FPGAs capture the digital signals and format and transmit the data. The logic in Figure~\ref{fig:IO-logic} samples the CLARO signals at 320\,Mbit/s using the deserialiser embedded in every input-output (I/O) logic block. The byte from the deserialiser is used to address a lookup table, which can be programmed to detect specific signal patterns arriving from the CLARO channel and to apply a time gate of 3.125 or 6.25\,ns. The delay control module in the I/O logic of the FPGA provides a stable and linear method to compensate the typically few nanoseconds channel-to-channel variations in ToA of the CLARO signals. The module can be programmed to delay each channel separately up to 3.2\,ns in steps of 100\,ps. The number of delay steps is obtained using a calibration procedure. The results of this procedure using a test charge at the CLARO input are shown in Figure~\ref{fig:FPGA-calib} for all channels of one FPGA and demonstrate the effectiveness to compensate the channel offsets. \par

The ToA and time-over-threshold (ToT) of the CLARO digital signal were characterised by shifting the FPGA sampling clock phase in steps of $\sim50$\,ps for a range of injected test charge amplitudes and threshold settings. Generally, the ToT information can be used to correct the time-walk effect of up to $\sim3$\,ns observed in Figure~\ref{fig:CLARO-toa-tot}. However, this correction is not foreseen in the Run~3 detector because the 3.125\,ns sampling bins in the FPGA limits the ToT resolution and the large channel-to-channel variation in ToT of 5 to 10\,ns (indicated using the error bars) would require a per-channel custom correction. Figure~\ref{fig:testpulse-timewalk} shows the average shape of the digital signal captured by the FPGA for a given test charge amplitude. Here, the width of the transition from zero to one on the digital signal quantifies the electronics jitter including the CLARO output, PCB layout and FPGA input. This width is less than 100\,ps at large input charges (black curve) and degrades to $\sim150$\,ps for small charges (red curve) closer to the threshold. 
Although the detector could be operated with a 3.125\,ns hardware time gate, the combined spread from time walk and the PV distribution (adding $\sim0.5$\,ns FWHM) may require a  6.25\,ns gate in practice. \par

\vspace{-0.2cm}
\section{Towards the future: picosecond RICH detector during HL-LHC Run 5}

\begin{figure}[t]
	\begin{center}
		\includegraphics[width=0.9\linewidth]{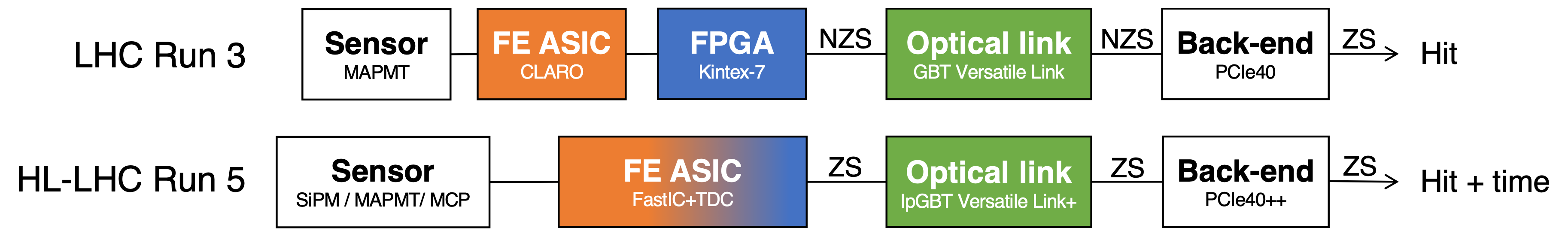}
	\end{center}
	\vspace{-0.7cm}
	\caption{\label{fig:electronics-overview}Overview of the electronic readout from Run~3 to Run~5. The novel FE ASIC will have added fast-timing capability and zero-suppression (ZS) to limit the bandwidth increase.}
	\vspace{-0.4cm}
\end{figure}

The increase in luminosity during HL-LHC Run~5 causes a challenging rise in particle multiplicity and hit occupancy in the detector. Amongst other improvements, a time resolution of better than 100\,ps is expected to maintain the high particle ID performance after Upgrade~II. The increased irradiation level dictates a shift from FPGAs to a more radiation-hard ASIC design as outlined in Figure~\ref{fig:electronics-overview}. The proposed design integrates analog and digital functionality into a single ASIC that is tightly integrated with the next-generation optical links. The data bandwidth will rise due to the higher hit occupancies and added timing information. The ASIC will therefore have zero suppression, time gating and constant-fraction discrimination. Additionally, compact electronics with low power consumption are required especially due to the high channel density and low operating temperatures for silicon photomultipliers (SiPMs). This novel ASIC is proposed to be introduced during LS3 to improve the particle ID in Run 4 and to transition to Upgrade~II. The FastIC is a 65-nm CMOS chip with a single-photon time resolution of $\sim25\,$ps~\cite{FastIC}. Its wide dynamic range and dual polarity would be suitable for MAPMTs during Run~4 and SiPMs during Run~5. The ‘FastIC+TDC’ custom adaptation will include many of the RICH requirements as input parameters and would be a highly suitable ASIC for the RICH upgrade programme. The alternative of a digital-only custom ASIC for Run~4 to replace the FPGA and read out CLAROs with $\sim390$\,ps time bins is also being studied.\par

\vspace{-0.2cm}
\section*{Conclusion}

The intrinsic time resolution of the RICH detectors of better than 10\,ps provides a strong motivation to develop a fast-timing photon detector module to improve the particle ID performance. Photon timing techniques and expected performances were presented. During LHC Run~3, a time gate between 3 and 6\,ns with channel-to-channel delay calibration will be implemented in the front-end FPGA and used to verify the simulation studies. During HL-LHC Run~4 and 5, a single ASIC is envisaged with a time resolution better than 100\,ps. The described RICH requirements are input to the design of the FastIC+TDC custom ASIC.\par

\vspace{-0.2cm}
\section*{References}

\bibliography{Keizer-TIPP2021}
\bibliographystyle{unsrt}

\end{document}